# *Radial Injection in Suspension High Velocity Oxy Fuel (S-HVOF) Thermal Spray of Graphene Nanoplatelets for Tribology*

*F. Venturi, T. Hussain\**

*Faculty of Engineering, University of Nottingham, Nottingham, NG7 2RD, UK*

*\*Tanvir.Hussain@nottingham.ac.uk ; +44 115 951 3795*

**Abstract**

Friction is a major issue in energy efficiency of any apparatus composed of moving mechanical parts, affecting durability and reliability. Graphene nanoplatelets (GNPs) are good candidates for reducing friction and wear, and suspension high velocity oxy-fuel (SHVOF) thermal spray is a promising technique for their scalable and fast deposition, but it can expose them to excessive heat. In this work, we explore radial injection of GNPs in SHVOF thermal spray as a means of reducing their interaction with the hot flame, while still allowing a high momentum transfer and effective deposition. Feedstock injection parameters, such as flowrate, injection angle and position, were studied using high-speed imaging and particles temperature and velocity monitoring at different flame powers using Accuraspray 4.0. Unlubricated ball-on-flat sliding wear tests against an alumina counterbody ball showed a friction coefficient reduction up to a factor 10 compared to the bare substrate, down to 0.07. The deposited layer of GNPs protects the underlying substrate by allowing low-friction dry sliding. A Transmission Electron Microscopy study showed GNPs preserved crystallinity after spray, and became amorphised and wrinkled upon wear. This study focused on GNPs but is relevant to other heat- and oxidation-sensitive materials such as polymers, nitrides and 2D materials.

# 1 Introduction

The unprecedented mechanical [1] and tribological [2] properties of graphene have attracted a wide range of interests in using it as a solid lubricant [3]. The lubricating effect favors a lowering of the coefficient of friction and a delay in damaging the lubricated surfaces. This can effectively improve the durability of moving mechanical parts, by reducing localized heating and subsequent wear. Some small-scale deposition techniques [4, 5] can be hardly employed for covering large areas with a considerable amount of graphene. Simple techniques like drop casting and airbrush spray [6] would allow the deposition; however, they do not provide a good bonding with the substrate. Conversely, spray techniques such as supersonic cold spray [7] proved suitable for large scale graphene coverage and enhanced the bonding with the substrate due to the high velocity at impact. An even better graphene-substrate adhesion could be reached with thermal spray, as it gives not only high kinetic energy but also a higher amount of heat to the particles. In particular, Suspension High Velocity Oxy Fuel (S-HVOF) thermal spray [8] is a good candidate for this task. This relatively new technique allows the injection and spray of suspension in HVOF instead of powders, thus allowing to handle finer particles, down to the micro-to-nano scale. Overall, this technique provides high acceleration and has a relatively low flame temperature compared to plasma spray. Graphene is known to be stable at high temperatures in an inert environment, however in air it starts degrading at around 250 °C, and a consistent mass loss can occur at around 500 °C, according to thermogravimetric analyses (TGA) [9]. The use of graphene nanoplatelets (GNPs) could be more suitable due to their higher stability. GNPs consist of a stack of 15-20 graphene layers and are characterized by a surface to volume ratio lower than single layer graphene, which improves their stability at high temperatures. According to TGA analyses in air they show a limited mass loss at 500 °C followed by a consistent mass loss only at around 700 °C [10]. GNPs retain the good mechanical properties of graphene and also provide the lubricating characteristics of lamellar solids [11]. GNPs have been already used in form of nanocomposites with alumina [12, 13] providing good wear response [14], thanks to the formation of a protective layer incorporating graphene.

Considering the use of S-HVOF thermal spray, care must be taken in the way the feedstock is injected into the flame, since at this stage GNPs degradation can occur. There are two main ways of injection in thermal spray: the axial injection, where the feedstock is injected directly into the combustion chamber, and the radial injection, where it is injected into the flame from the side, outside the gun. Radial injection is the ordinary route for injection in plasma spray, where it has been shown that changing feedstock injection position and angle can tailor the amount of interaction with the flame in order to spray very different materials, from ceramics to polymers and their composites [15]. However, radial injection has been rarely employed in HVOF thermal spray, but has been proven useful for depositing materials while avoiding temperature dependent phenomena, such as anatase to rutile transformation in titanium oxide [16]. As a downside, radial injection in HVOF thermal spray can lead to the deposition of only partially melted or unmelted $TiO_2$ particles [17]. In fact, axial injection leads to a long residence time of the feedstock in the high-temperature part of the jet, whereas radial injection leads generally to a lower degree of feedstock-flame interaction. For the spray of GNP, the axial injection and might lead to their degradation, whereas radial injection gives a smaller amount of heat, tunable by varying the injection parameters, yielding a lower degree of degradation. The injection axial position and direction for instance can have a noticeable effect on flame-GNP heat transfer and have to be carefully chosen. The interaction with the hot jet must be a trade-off between providing high kinetic energy and at the same time hindering mechanical and thermal degradation of GNPs.

In this work, different radial injection parameters were explored, such as feedstock flowrate, injection angle and axial position, and their effectiveness in penetrating the jet and allowing water carrier vaporization was analyzed. These parameters were studied at different flame powers as these can strongly influence the interaction dynamics. In addition, GNP in-flight temperature and velocity was measured at different conditions, yielding information on the heat and momentum exchange with the jet. The optimized set of parameters was then used to deposit a GNP film on stainless steel substrate. The GNP film was characterized by studying its morphology and its tribology performance compared to a reference stainless steel substrate. The morphological and structural state of GNPs before and after wear were studied with Transmission Electron Microscopy (TEM).

## 2 Materials and Methods

### 2.1 GNPs Suspension Preparation

A deionized water-based suspension was prepared using 1 wt. % GNPs (abcr GmbH, Germany), with nominal 5 µm average width and 5-8 nm thickness (15 - 20 graphene layers), and 0.01 wt. % sodium dodecyl sulfate surfactant (Sigma-Aldrich Company Ltd., United Kingdom) to hinder particles agglomeration. The suspension was stirred using a FB-505 sonic dismembrator (Fisher scientific, United Kingdom), at 20 kHz with 2 s pulse every 5 s, for 3 hours and again for 30 minutes prior to spraying. The stability of the suspension, assessed through a zeta potential measurement using a Zetasizer nano-ZS (Malvern Panalytical, United Kingdom), led to a value -40.2 mV, highlighting its high stability.

### 2.2 S-HVOF Thermal Spray

S-HVOF thermal spray was carried out with a commercial GTV TopGun (22 mm combustion chamber, 135 mm barrel, 8 mm internal diameter), using oxygen as working medium and hydrogen as fuel. The duct for axial injection was blocked with a brass plug. Initial tests were carried out for studying the different injection parameters using high-speed imaging and particles temperature and velocity monitoring. These tests were done at different gas flowrates to investigate different flame powers, obtained from combustion calculations, as shown in Table 1.

**Table 1: Fuel flowrates.** Various oxygen and hydrogen flowrates used for obtaining different flame powers.

| Flame power (kW) | Oxygen flowrate (l/min) | Hydrogen flowrate (l/min) |
|---|---|---|
| 25 | 78 | 182 |
| 50 | 151 | 354 |
| 75 | 227 | 533 |

The choice of these values was aiming at exploring a wide range of S-HVOF thermal spray operating regimes, from a very low subsonic 25 kW flame power, ensuring little damage to GNPs, a moderate, subsonic 50 kW flame power, and a high, supersonic 75 kW flame power, providing high momentum but possibly damaging GNPs. The choice of non-stoichiometric ratio of oxygen and hydrogen ensures a reducing environment which hinders GNP oxidation.

GNPs were deposited on 304 stainless steel substrates (nominal composition: 18% Cr, 8% Ni, 2% Mn, 0.15 % 0.08% C, 0.045% P, 0.03% S, 0.75% Si, 0.1% N, 70.845% Fe, all in wt. %) measuring 60×25×2 mm, polished down to a 1 µm diamond finish (grinding disc grit size P240, P400, P800 and P1200, and diamond pad polishing at 6 µm and 1 µm). The substrates were mounted on an air-cooled carousel with a diameter of 26 cm, rotating at 73 RPM. The deposition consisted of 2 consecutive spray runs of 20 passes each. The spray gun was pointing perpendicularly to the carousel axis and moved along the z axis at a speed of 10 mm/s.

The radial GNP suspension injection was obtained using a custom attachment for the S-HVOF thermal spray gun as shown in Fig. 1a. The suspension has been injected using a commercial XMW 4001 T8 1/4" air-atomizing nozzle (PNR, United Kingdom) used without air, actually corresponding to a pure injector with internal duct diameter of 450 µm. The custom atomizer holder attachment allowed for the choice of the injection direction as well as the axial and radial location of the atomizer. By varying the air pressure in the feedstock chamber, the suspension feed rate is controlled, and measured using an ES-Flow low flow ultrasonic flowmeter (Bronkhorst Ltd., United Kingdom).

### 2.3 In-flight Measurements

High-speed imaging was performed with a V12 High Speed camera (Phantom, USA). The image resolution with this setup is 10 px/mm and an exposure time of 1 ms was used. An even back illumination was obtained by illuminating a white background. The high contrast was provided by the illumination discrepancy between the bright background and the darker spray apparatus and injected feedstock in the foreground. A glass screen prevents damage to the camera while screening the CCD sensor from excessive UV light from the HVOF flame, which is therefore barely visible. This setup allows for proper imaging of the feedstock injection into the flame, with following displacement and eventual vaporization, as shown in Figure 1b.

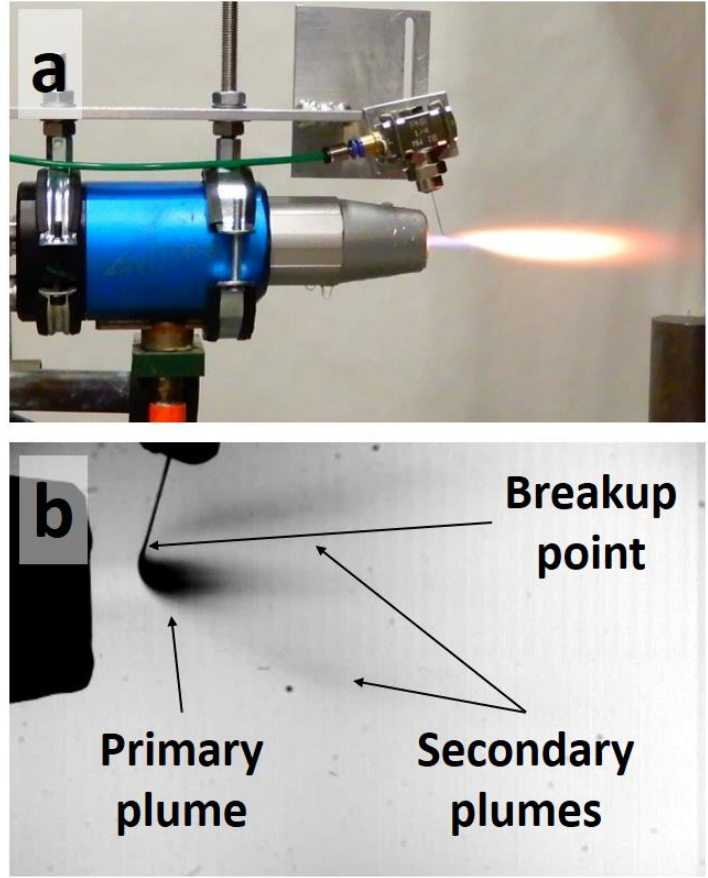

**Figure 1 – Experimental set-up and high-speed imaging.** a) HVOF gun with the custom attachment for radial injection using an atomizer. The system is in operation: the initial jet can be seen, whose color and width are changed by the injection of GNPs. b) example of a high-speed image radial injection of water. The breakup point, primary and secondary plumes can be observed.

Here, also the water column breakup and primary and secondary plumes can be identified. The water column breakup point is where the droplets start forming; in the image the droplets appear as a solid plume due to the length of the exposure time. The primary plume is where most of the feedstock directs; in some cases, mainly due to rebounding by the flame or due to its overreaching, part of the feedstock is displaced towards other directions, forming secondary plumes which cause overspray. The feedstock used for high speed imaging was deionized water. Since the GNP solid load is 1 wt. % only, the dynamics observed for water are comparable to the actual feedstock used for deposition and other low concentration feedstock in general, as their liquid columns have comparable momenta. The graphs showing the feedstock penetration into the jet have been carried out tracing the upstream side of the liquid column [18] using ImageJ (NIH, USA); the images were first converted to black and white by applying the same threshold of 90% to each image.

In-flight particles velocity and temperature measurements were carried out using Accuraspray 4.0 (Tecnar, Canada) pointing at a stand-off distance of 100 mm. The signal amplification factor and exposure time settings were chosen different for different flame powers, but were comprised in the range of 24 – 28 times and 16 – 41 ms, respectively. These values were chosen to provide a good signal to noise ratio and to avoid saturation of the sensors. The response time was set to 5 s. Series of 30 measurements were acquired over a time frame of 30 s, then averaged to give the resulting values. The standard deviation σ of the series of measurements acquired, which indicates their precision, was calculated for temperature and velocity measurements to be $\sigma_T < 2\%$ and $\sigma_v < 3\%$. It should be noted that fluctuations in the combustion gases and feedstock flowrate can lead to particles temperature and velocity fluctuations; therefore, these standard deviations not only represent a measurement precision but they are also an actual distribution of values. Also, the accuracy of the measurement as stated by the manufacturer is 3% both for temperature and for velocity measurements. As an upper boundary for the measurement error, the accuracy error was associated to the experimental values presented in this work. Accuraspray 4.0 is routinely used for powder HVOF and suspension plasma and not much work was been done on S-HVOF. However, GNPs are a suitable system to study as they have a high specific surface, leading to good heat radiative emission. Also, as velocity measurements are carried out on the overall flame and not on the single fine particles, the measurements of suspension in this system doesn't cause detection issues related to the size of the particles.

### 2.4 Tribology

Tribology measurements were carried out with a ball on disc tribometer (Ducom instruments, The Netherlands) using an $Al_2O_3$ spherical counterbody with 6 mm diameter. The measurements were carried out in ambient conditions. Friction coefficient measurements were done at 2 N load, along a circular path of 10 mm diameter at an angular velocity of 60 RPM. The total wear distance was 31.415 m over a time of 16 min 50 s corresponding to a total of 1000 wear cycles. The obtained data were frequency-filtered using Matlab (The Mathworks inc., USA) with a 45 dB bandstop filter between 55 and 75 mHz with 0.85 steepness. The cut frequency, 65 mHz, corresponds to the beating frequency from the revolution frequency of the wear test, which causes small oscillatory effects in the frictional force readings due to the non-perfect planarity of the sample [19], and the sampling frequency. The filtering removes this contribution allowing a greater insight into the tribology properties of the system.

### 2.5 Samples Characterization

The samples morphology was studied with scanning electron microscopy (SEM) using a XL30 microscope (FEI, The Netherlands) with tungsten emitting filament at 5 kV accelerating voltage. No sample preparation was needed as both GNPs and stainless steel allow sufficient electrical conduction.

TEM analyses were carried out using a JEM2100+ microscope (JEOL, Japan) equipped with $LaB_6$ electron source and operated at 80 kV accelerating voltage to avoid damaging the sample with the electron beam. Electron Energy Loss Spectroscopy (EELS) measurements were obtained using an Enfinium detector (Gatan, USA); a power law-type background was subtracted from the spectrum to extract the signal. The TEM sample preparation consisted of scratching the GNP sample surface with a craft knife in order to collect a sample of GNPs. The craft knife was previously cleaned with industrial methylated spirit (IMS) to minimize contamination. This operation was carried out both in the as-sprayed regions of the sample and inside the wear track of the wear test, where the GNP protective layer was present. The scratched sample was transferred on a TEM holey carbon copper grid.

## 3 Results and Discussion

### 3.1 Radial Injection Parameters Window and High-speed Imaging

The use of the customized radial injection setup required an optimization process aiming at choosing the combination of parameters which maximize feedstock penetration in the jet, without excessively increasing its residence time, and preventing overspray i.e. no deposition on the target substrate. Other important effects to be assessed were the primary plume length and width, and eventual secondary plumes. Three parameters were changed in this study: the feedstock feed rate, the injection angle and the injection position, i.e. the distance of the injection from the gun exit. In this high-speed imaging study the feedstock consisted of water only.

#### 3.1.1 Effect of distance

Preliminary studies on the injection at different distances from gun exit were done at the following parameters: 200 ml/min flowrate, perpendicular injection and 50 kW flame power. This first test confirmed that a short distance between injection position and gun exit favored proper carrier water vaporization, as shown in Fig. 2.

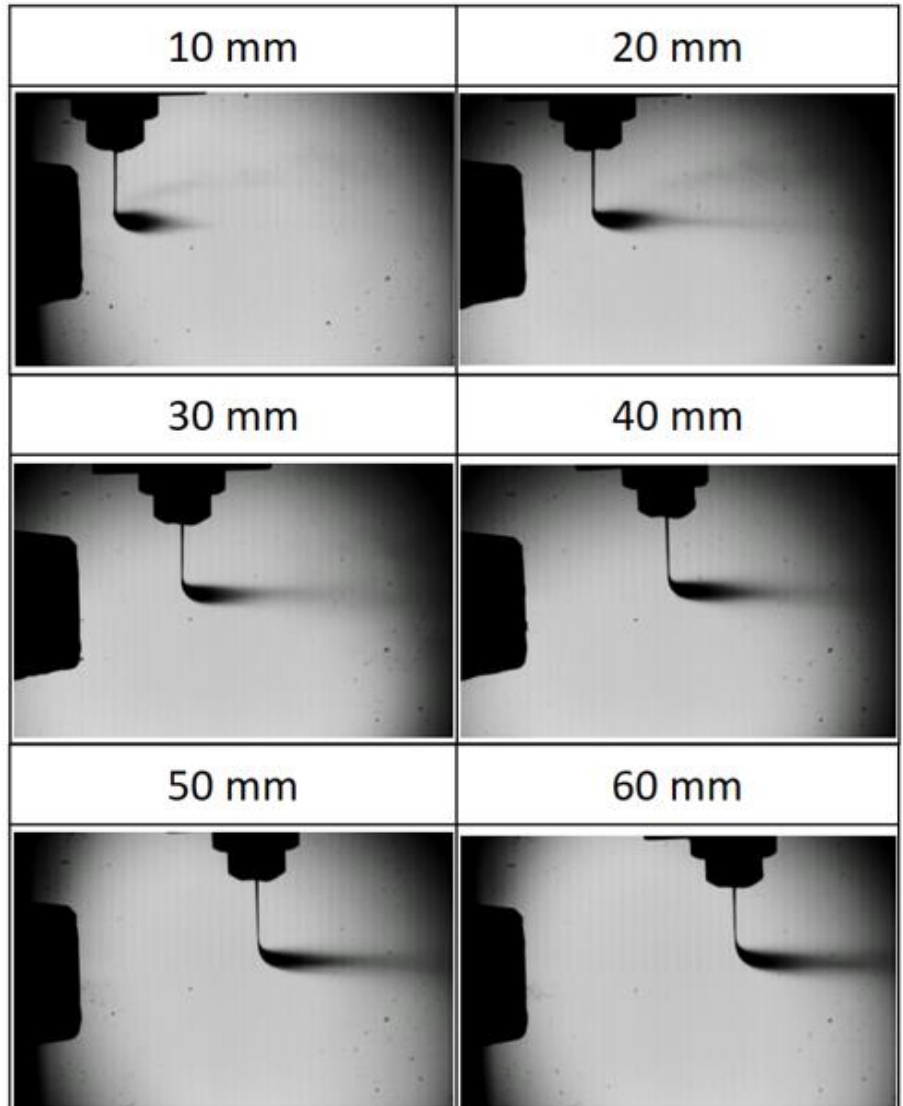

**Figure 2 – High-speed imaging: varying injection position.** High-speed images of the radially injected feedstock entering the HVOF jet. The feedstock was injected perpendicularly from the top, the jet originates from the gun at the left side and is barely visible. The injection distance from gun exit was studied, varying from 10 mm to 60 mm in 10 mm steps, showing lower rates of vaporization as the distance increases.

The vaporization efficiency can be revealed by looking at the length of the plume the injected feedstock traces in the jet. At 10 mm injection distance a full vaporization is obtained, as the plume completely disappears before reaching the right hand side of the image. A small secondary plume can also be seen heading upwards in the 10 mm and 20 mm case, due to the higher local velocity of the flame at those distances. At all longer injection distances, the plume length is gradually increasing and can be seen all the way through the image up to the right border, indicating an incomplete vaporization, which would lead the liquid part of the feedstock to unwantedly reach the substrate. The injection distance was therefore chosen and fixed to 10 mm from this point on. Also, since the feedstock flux regime is laminar within short distance and the injection comes from above the jet, the radial injection distance from the jet axis does not affect the penetration dynamics except for a small factor due to gravity which can be neglected. This distance was then fixed to 20 mm.

### 3.1.2 Effect of flowrate

Once completed the trials for injection distance, the feedstock injection flowrate and the injection angle were studied. The high-speed images taken at the various injection conditions are presented in Fig. 3 for the 25 kW, 50 kW and 75 kW case. Here, a series of feedstock feed rates, from 50 to 300 ml/min in 50 ml/min steps, are presented. Also, three injection angles are explored: 15° upstream, perpendicular to jet and 15° downstream.

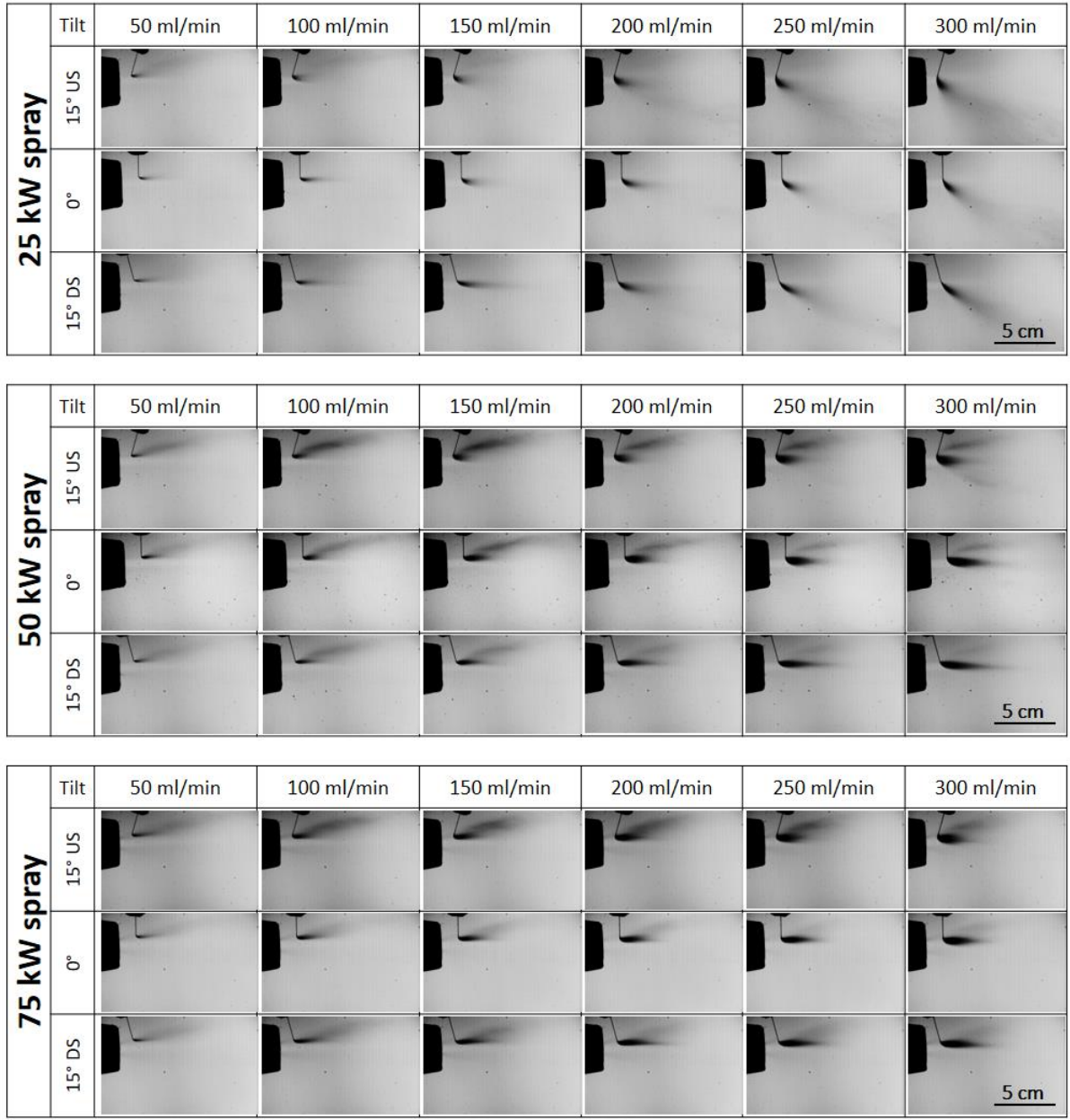

**Figure 3 - High-speed imaging: varying flame power, flowrate and injection angle.** The different charts present results for the flame powers of 25, 50 and 75 kW. In each chart, the three rows present results of 15° upstream (US), perpendicular and 15° downstream (DS) injection, and the six columns present injection at different flowrates, from 50 to 300 ml/min.

This first set of images allows to draw some qualitative conclusions. The feedstock penetration in the jet appears higher at higher flowrates and lower flame power. This first observation is as predicted since increasing the flowrate while keeping the duct diameter constant yields a liquid column with higher momentum, and a lower power jet can be penetrated more easily as its pressure, velocity and temperature are lower. The jet is fully overtaken only at 25 kW with flowrates > 200 ml/min. This condition leads to overspray and is not suitable for deposition.

Concerning the length of the primary plume, another observation is its greater length at higher flowrates, as a higher volume of liquid is available per unit time. More interestingly, the plume at a given flowrate is longer for downstream injection and shorter for upstream injection, with the perpendicular case in between. This is because in the upstream case a backward momentum has to be overcome and reverted by the jet, concentrating vaporization into a smaller region and leading to slightly longer residence time into the jet. Additionally, the plume for upstream injections not only shortens but also widens, possibly leading to a more sparse spray or overspray. The opposite is true for downstream injection. Also, the widening of the plume due to upstream injection leads to a partially higher penetration as can be seen in Fig. 3 at 50 kW, 300 ml/min, where only the upstream case shows a secondary plume below the primary plume. The term partially refers to the fact that not all the plume overreaches the jet but only a small part of it due to the widening given by upstream injection. Also, here at all injection angles a secondary upwards plume is formed. The formation of upwards secondary plumes is overall particularly evident in the upstream series. These secondary plumes are minimum at moderate flowrates with downstream injection, which is a favorable condition since all kind of secondary plumes yield overspray.

### 3.1.3 Water column breakup and injection angle

A quantitative analysis was carried out involving the set of images in Fig. 3. The quantities investigated were the distance of water column breakup from jet axis at different flowrates, flame powers and injection angles, as a quality factor for the degree of penetration achieved. This is an important parameter because, as the water column breaks and the droplets become smaller, the surface available for heat exchange with the flame increases, nonlinearly increasing the vaporization rate. The water column breakup distance values at different flowrates and flame powers with perpendicular injection are shown in Fig. 4a, whereas those for different injection angles at different flame powers at a fixed flowrate of 200 ml/min are shown in Fig. 4b.

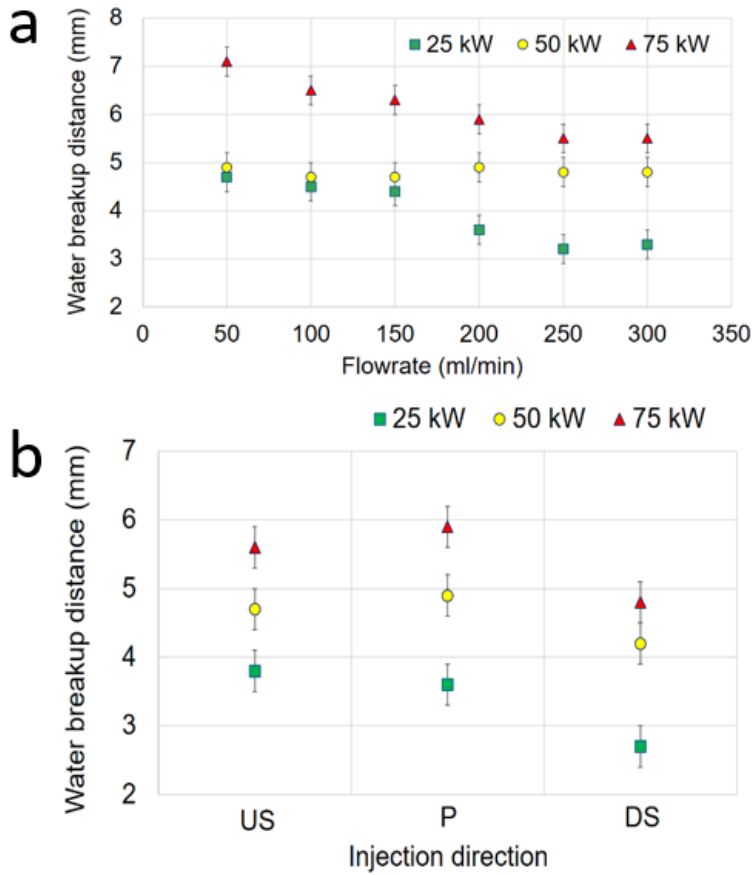

**Figure 4 – Feedstock liquid column breakup.** a) Liquid column breakup distance from jet axis at different flame powers and injection flowrates and b) at different flame powers at 200 ml/min flowrate for 15° upstream (US), perpendicular (P) and 15° downstream (DS) injection angles.

Overall, the breakup distances from the jet axis are all in the 3-7 mm range, with shorter distances at lower flame power. This is expected, as a lower power flame is characterized by lower expansion shock, and lower velocity, pressure and temperature, therefore the injected feedstock is less influenced by it. Considering now Fig.4a, a higher feedstock flowrate, hence a higher injection momentum, provides a breakup which is closer to the jet axis. It is however noticeable how the 50 kW case was not affected by the change in flowrate. The subsonic to supersonic threshold is around 60 kW flame power, therefore in the 50 kW case, where there is a strong subsonic flame, the breakup was not much affected by the change in flowrate, while the 25 kW case, with a weak subsonic flame, definitely it is affected. In the 75 kW case the flame is supersonic, therefore there is as an additional effect the sudden expansion with the formation of shock diamonds, typical of a supersonic flame. In this case the breakup distance is affected by the flowrate, especially at lower flowrates where the breakup distance highly increases, but overall the breakup distance is larger than at other flame powers.

The other interesting analysis, shown in Fig.4b, investigates the water column breakup at different injection angles: 15° upstream, perpendicular and 15° downstream injection. Upstream and perpendicular injection directions yield similar results in terms of breakup distance, whereas for downstream injection the breakup distance decreases at all flame powers. In downstream injection, the component of the injected feedstock momentum which is parallel to the jet direction allows for a smoother breakup, which ultimately takes place at a later stage. This observation suggests that a downstream injection can be more suitable for the feedstock to reach the jet axis and to form a narrower plume.

### 3.1.4 Feedstock trajectories

A quantitative insight into the penetration of the feedstock into the jet is given by the graphs in Fig. 5, where the upfront side of the injected water column is plotted.

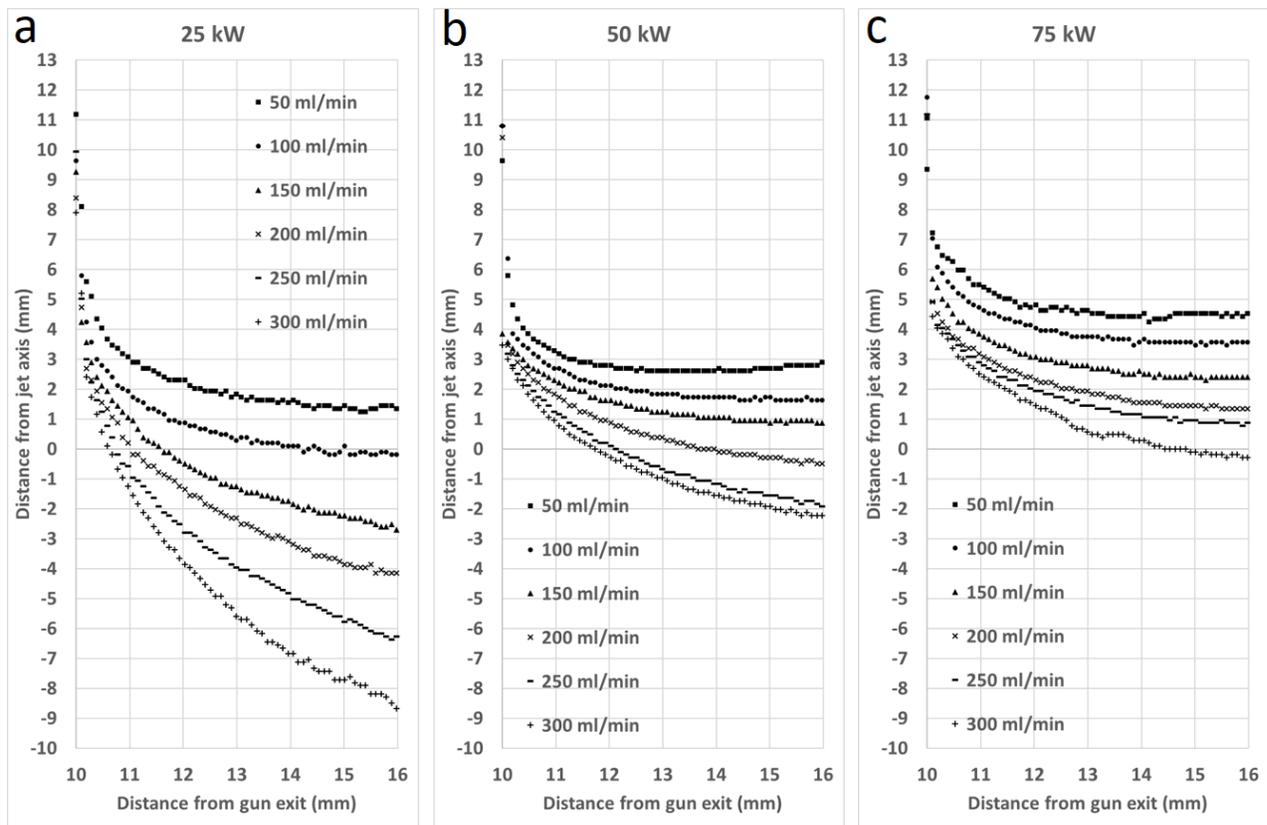

**Figure 5 - Feedstock injection trajectories.** Liquid trajectories (upstream side - perpendicularly injected) through the jet at different flowrates for a 25 kW (a), 50 kW (b) and 75 kW (c) flame. A higher penetration is achieved at higher flowrates and lower flame power.

A common feature is the progressively higher amount of penetration as the flowrate increases. However, it can be seen how this effect is more pronounced in the 25 kW case, and far smaller in the 50 kW and 75 kW case. This happens because in these two latter cases the flame so strong that its core can be barely overreached even at the higher flowrates. The main difference between the 50 kW and 75 kW case are not the trajectories themselves but the overall penetration which is 2 mm less in the 75 kW case. Conversely, in the 25 kW case the trajectories follow a different kind of path. In particular, those that overreach the flame (flowrate > 100 ml/min), once done so travel according to their inertia and gravity instead of following the jet trajectory. From these sets of trajectories it can be concluded that flowrate values which yield suitable penetration are 50-100 ml/min at 25 kW, 150-200 ml/min at 50 kW and 250-300 ml/min at 75 kW. All the previous considerations were done on a water-only feedstock since the dynamics are also representative of a low weight load feedstock. At a given flowrate, from a mechanical point of view and disregarding viscosity, the increase in solid load will increase the momentum and deviate from this analysis yielding a higher degree of penetration.

### 3.2 In-flight Temperature and Velocity Measurements

In this section, results on the actual GNP 1 wt. % feedstock are presented, for which measurements of temperature and velocity of particles have been carried out. A set of measurements that study how temperature and velocity of GNP vary according to changing flame power, flowrate and injection angle, are presented in Fig. 6.

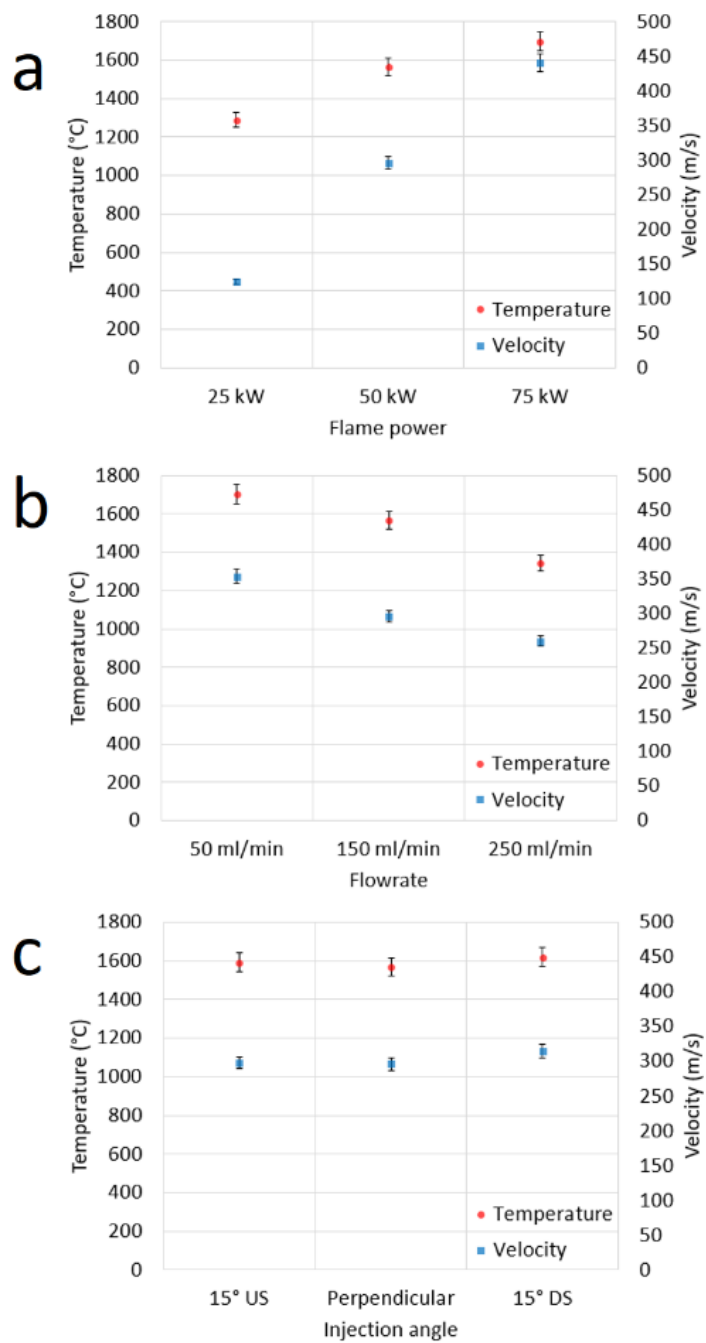

**Figure 6 – GNP in-flight temperature and velocity measurements.** a) Measurements taken varying flame power and keeping flowrate (150 ml/min) and injection angle (perpendicular) constant. b) Measurements taken varying flowrate and keeping flame power (50 kW) and injection angle (perpendicular) constant. c) Measurements taken varying injection angle and keeping flame power (50 kW) and flowrate (150 ml/min) constant.

A clear trend can be seen in Fig. 6a where, at a fixed flowrate of 150 ml/min, both temperature and velocity of GNPs increase with increasing flame power, reaching T = (1698 ± 51) °C and v = (441 ± 13) m/s with a 75 kW flame power. This is an interesting result as it shows that, despite penetration at a given flowrate is lower for higher flame power, the GNP still manage to be accelerated and heated more. Due to the nature of these measurements, the readings are taken at the center of the primary plume and don't take into account the eventual secondary plumes. Therefore, it should be taken into account that even if higher values are reached at 75 kW, these could refer to a smaller fraction of particles compared to the 50 kW and 25 kW case. Additional considerations on these measured values can be made upon comparison with flame temperature and velocity values simulated using the eddy dissipation concept model from a recently published work on this same HVOF setup [20]. According to these simulations, the 75 kW flame at 100 mm stand-off distance is characterized by a gas temperature and velocity around 1800 °C and 700 m/s respectively, which are reasonable upper boundaries to the values measured in this work. Also according to these simulations, a radially injected flowrate of 150 ml/min only slightly reduces the gas T and v values, with relevant reductions only at flowrates > 200 ml/min, therefore explaining the lower values measured here. It should be noted that most particles employed in thermal spray have a shape which can be reasonably considered a sphere, whereas GNPs aspect ratio is very much dissimilar to a sphere. This has aerodynamic and thermodynamic implications, in fact, GNPs tend to exchange heat and momentum much quicker with the medium they are immersed in, and this means they tend to follow the gas velocity and temperature more than spherical particles. Also, because of their lightweight and low density, even at high velocity their momentum is relatively small and they tend to slow down quickly as the flame does so.

Figure 6b shows, for a 50 kW flame, how an increase of feedstock flowrate yields lower GNP velocity and temperature. The decrease is consistent, of about 400 °C and 100 m/s starting from 50 ml/min to 250 ml/min, in which conditions values down to T = (1342 ± 40) °C and v = (260 ± 8) m/s are reached. Due to the very low solid load of the feedstock, this behavior is mainly due to the water fraction and can be understood accordingly. The mechanical effect of injecting more water is the increase of mass that needs to be accelerated by the jet which results in a lower particle velocity. The thermodynamic effect is due to the amount of water that has to be vaporized. An increase of 100 ml/min of water injected into the flame requires an additional 3.76 kW to achieve complete vaporization. It is noteworthy how the effect of increasing penetration due to increasing flowrate, as shown in Fig. 5b, which provides a better heat and momentum transfer, is overwhelmed by this other effect due to the amount of water that has to be heated and accelerated. A more efficient exploitation of the higher penetration reached by increasing flowrate would be achieved by using a feedstock with higher solid load or with a liquid carrier having lower vaporization enthalpy such as ethanol. Another route to circumvent this issue would be using a smaller injector duct, which at a given flowrate yields higher feedstock velocity.

The third parameter investigated, the injection angle, is presented in Fig. 6c for a 50 kW flame and 150 ml/min flowrate, and shows very little difference between the three cases. In particular, upstream and perpendicular injection are comparable within the experimental error. Their major difference which emerged from Fig. 3 was the plume width, wider for upstream injection, narrower for downstream and with the perpendicular case in between. This trend doesn't seem to affect the GNP velocity and temperature at all; however, the downstream case here shows a small increase in both velocity and temperature of the particles. This trend resembles the one presented in Fig. 4b, where downstream injection provides a breakup distance closer to the jet axis. Therefore, a downstream injection appears to be beneficial for better reaching the core of the jet and exchanging heat and momentum.

In summary, these analyses show that the feedstock interaction with a 25 kW flame is very much affected by the injection parameters, with the risk of overreaching the flame at high flowrates, causing overspray. Alongside, the feedstock interaction with a 75 kW flame is barely affected by varying the injection parameters, with the jet core inaccessible even at high flowrates, preventing the feedstock from exploiting the full available heat and acceleration. A 50 kW flame, with in-between characteristics, was chosen as the most favorable for this purpose. The overall criteria for the choice of the best parameters for GNP feedstock involve good penetration, short and narrow plume, and minimal secondary plume. The set of injection parameters chosen for GNP deposition was then the following: the injection angle was chosen pointing 15° downstream, and the suspension feed rate was selected to be 170 ml/min at $2 \times 10^5$ Pa pressure.

### 3.3 GNP Thin Film Deposition

With the chosen set of injection and spray parameters (50 kW flame power, 15° downstream injection direction, 170 ml/min feedstock flowrate), a deposition was done on 1 µm-polished stainless steel substrates at a stand-off distance of 300 mm from the gun exit. The previous velocity and temperature study was carried out at 100 mm stand-off distance and no measurement was possible at 300 mm stand-off distance because the temperature of the particles there was lower than the lowest temperature detectable by Accuraspray 4.0 (1000 °C). In the additional 200 mm, the temperature and velocity of the GNPs are expected to decrease to a great extent, due to their light weight and high aspect ratio (their width is $10^3$ orders of magnitude larger than their thickness). Therefore, their heat exchange with the environment through conduction and radiation will be fast, the momentum they have gained will be reduced quickly and they will tend to follow the air stream. The stand-off distances used in S-HVOF thermal spray are normally much shorter than those used in this work, at around 85 mm. Such a short stand-off distance proves unsuitable for GNP spray since the hot jet (1850 °C at 85 mm for a 75 kW flame [20]), as it sweeps the substrate surface, transfers an amount of heat that accumulates on the sample and which GNPs are not able to withstand and may degrade. Moreover, the combustion jet from the S-HVOF thermal spray gun is capable of mechanically removing the loosely bonded GNPs from the substrate surface, hence lowering the deposition efficiency. Increasing the stand-off distance too much on the other hand would be beneficial for the low amount of heat transferred to the substrate, but the momentum retained by the GNPs at the impact would not be enough for proper adhesion. For these reasons, a stand-off distance of 300 mm was chosen as a compromise between these possible issues. The SEM top surface image of the as-deposited sample is shown in Fig. 7. The surface of the sample is almost fully covered by GNPs. Statistical analyses over 0.3 mm$^2$ yielded a coverage value (92 ± 2) %.

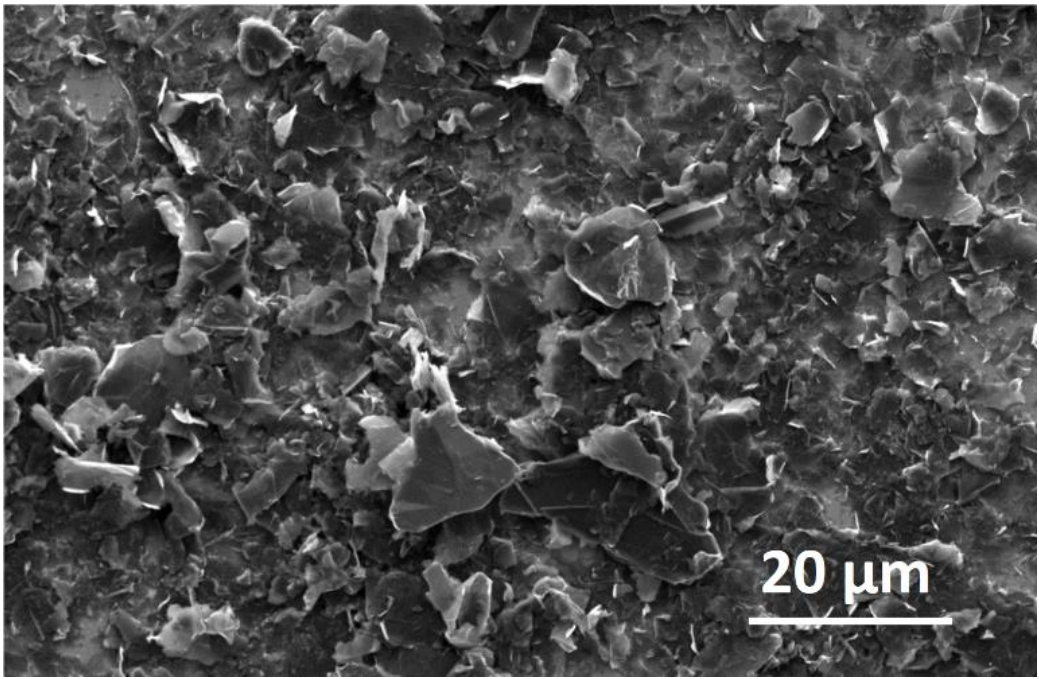

**Figure 7 - SEM of deposited GNPs.** SEM secondary electrons image of top surface of GNP-covered samples sprayed at 300 mm stand-off distance. The light grey contrast was given by the substrate. The presence of graphene hinders the signal of secondary electrons generated by the substrate [21], leading to darker contrast as the graphene thickness increases. The very bright contrast was given by GNPs which are not well-attached to the underlying material and thus charge under the electron beam.

### 3.4 Tribology Tests

Friction coefficient measurements were carried out both on the GNP sample and on a reference stainless steel substrate for comparison. The results of this test are shown in the graph of Fig. 8a.

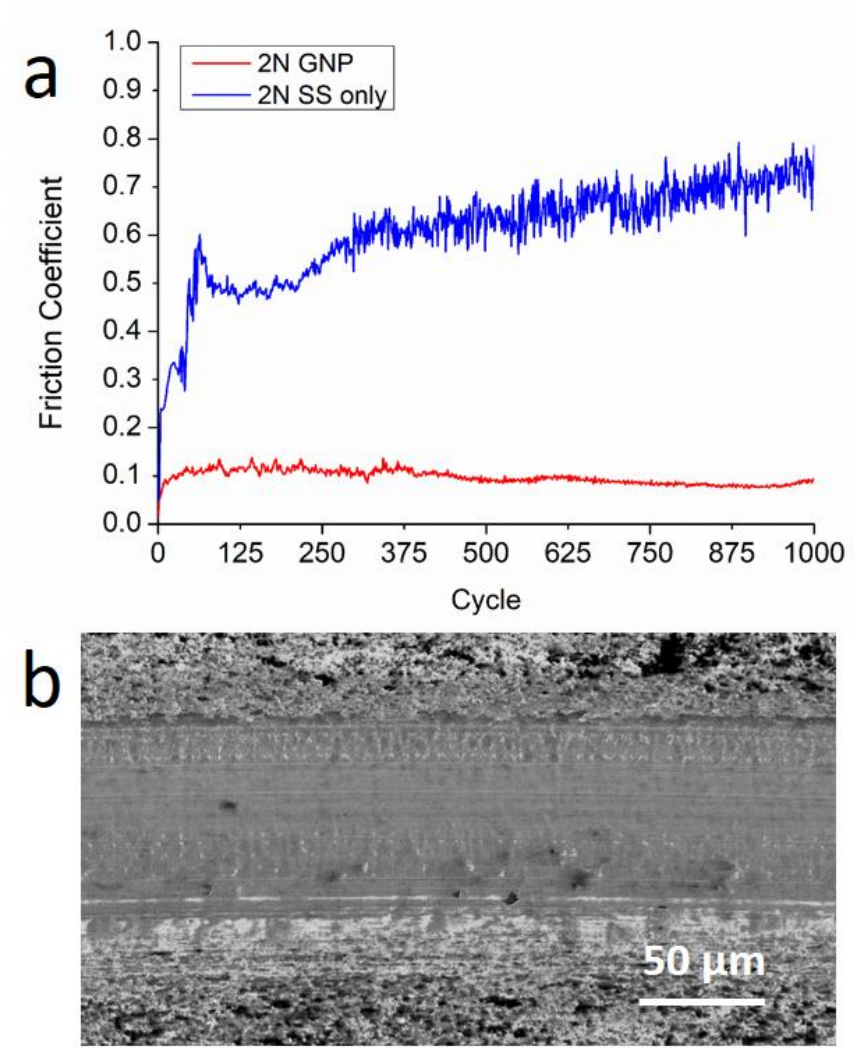

**Figure 8 – Tribology tests.** a) Coefficient of friction measurement at 2 N load over 1000 cycles for GNP on stainless steel (SS) sample and SS only sample. A reduction of the coefficient of friction up to 10 times is measured. b) Wear track showing a packed GNP layer which is formed upon wear, with a central smooth area and wrinkles at the sides.

The main striking feature is the dramatic decrease in coefficient of friction of the GNP sample with respect to the stainless steel only substrate. This decrease is up to one order of magnitude, reaching the lowest value of 0.07 after around 900 cycles. Such a small coefficient of friction can provide low energy dissipation through heating and prevent the underlying substrate from wear. It can be seen how friction coefficient slightly decreases throughout the 1000 cycles, and also tends to be more stable. This effect is due to the fraction of GNPs that are initially loosely bond or not parallel to the substrate. As the wear tests proceeds, these GNPs are progressively settled, oriented and packed in forming a smooth GNP layer. This later can be observed in the SEM image of Fig. 8b, where the wear track after 1000 cycles is shown. The wear track consists of this thin layer of packed GNPs, with a smooth central part and corrugated sides. This thin GNP layer allows a smooth sliding of the counterbody ball, as the graphene layers slide on each other providing a very low friction. As long as the GNP layer is preserved, the underlying substrate is protected from wear. This process leads to some degradation of the GNPs, as becomes evident from the corrugated wear track sides, and will eventually lead to a removal of the thin GNP film and a disruption of the wear protection. A deeper insight on the morphology and structure of the GNPs before and after spray is needed to understand how and why the low-friction characteristics degrade.

### 3.5 TEM Analyses

#### 3.5.1 As-sprayed sample

To better understand the GNP degradation upon wear testing, two samples were analyzed using TEM: as-sprayed GNPs and GNPs coming from the wear track after the end of the wear test. This analysis aims at studying how the structure and the morphology of GNPs is affected by thermal spray and wear. A TEM image of the as-sprayed GNP sample is shown in Fig. 9a.

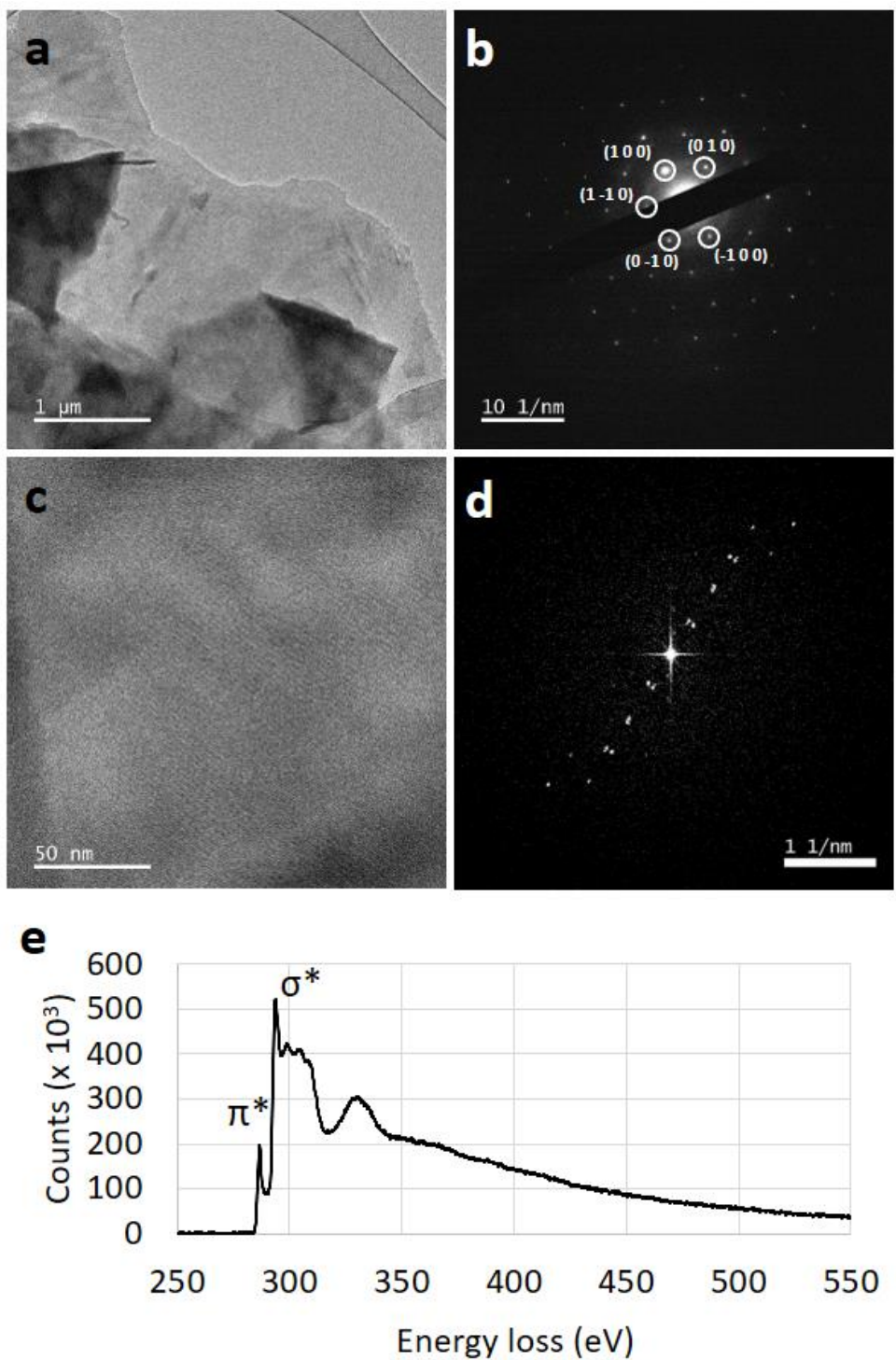

**Figure 9 – TEM of as-sprayed sample.** a) As sprayed GNP TEM image. b) Indexed SAED diffraction pattern from GNP in (a), showing the [001] zone axis. c) High resolution TEM image of another as-sprayed GNP. d) FFT of the HR image in (c) from which the lattice constant a = 2.46 Å is measured. E) TEM EELS spectrum from the particle in (a) centered on the carbon K-edge showing a typical graphite fingerprint with sharp π and σ peaks from the hybridized sp2 orbitals. No oxygen signal could be detected at its K-edge (532 eV).

Here, in the bottom left half of the image, one GNP particle is shown (the structure at the top right is part of the holey carbon copper grid support). This is a representative example of the kind of GNP found in this as-sprayed sample. The thinnest part, appearing light grey, corresponds to few layers of graphene, whereas the thicker, darker areas are formed by a stack of a higher number of graphene layers. As-sprayed GNPs are characterized by sharp edges and a typical crystalline structure. The selected area electron diffraction (SAED) shown in Fig. 9b, originated by the light GNP in Fig. 9a, highlights its crystallinity. This is a typical single crystal graphite diffraction showing the hexagonal arrangement of carbon atoms in the graphene layers. Due to their width, GNP tend to be preferentially oriented parallel to the substrate. In fact, the diffraction pattern shows the [001] zone axis, which means, the electron beam is impinging perpendicularly on the graphene layers. The indexing in the figure shows which family of planes gave rise to each reflection. This diffraction pattern proves the GNP structure is preserved upon S-HVOF thermal spray. Additional insight in the GNP structure is given by the High Resolution (HR) TEM image of another GNP particle presented in Fig. 9c. Here the high magnification allows to visualize intensity fringes originated by the atoms in the GNP. The main set of fringes is the one traversing the image from the top left to the bottom right. This set of fringes can yield additional information on the GNP structure by operating a Fourier transform of it. A Fast Fourier Transform (FFT) of Fig. 9c is shown in Fig. 9d. Here, a main set of periodicities is shown along the direction perpendicular to the fringes in Fig. 9c, as per Fourier theorem. This periodicity contains the information of one of the lattice constants of graphene, which is a = 2.46 Å as expected, which is the inverse of the distance of the first reflections from the center. The double points present indicate a stack misalignment between graphene layers, as also reported in [22]. A final spectroscopic analysis was carried out with EELS. With this technique it is possible to reveal the presence of elements as well as to study their chemical bonding. The EELS spectrum taken from the particle in Fig. 9a is presented in Fig. 9e. This is a high-loss spectrum centered on the Carbon K-edge, and shows a fine structure that is determined by the σ→σ* and π→π* transitions, as expected for $sp^2$ hybridized orbitals [23]. This fine structure is typical of crystalline graphite and, in our case, of GNP. This fine structure is very different from that expected by graphene oxide [24] and also no peak was detected at 532 eV, which id the Oxygen K-edge. Therefore, GNPs undergo minimal oxidation when sprayed and this is mostly limited to their surface, not affecting their inner structure. The temperature the GNP experience is very high compared to the one where they are stable in air. Therefore, they survive mainly thanks to the very short time they spend in the flame, of the order of milliseconds, and thanks to the oxygen-depleted flame environment that was chosen for this work. Overall, GNP that underwent radial injection and S-HVOF thermal spray are properly deposited and their crystallinity and composition is preserved.

### 3.5.2 Sample after wear test

TEM images of GNPs extracted from the wear track are presented in Fig. 10.

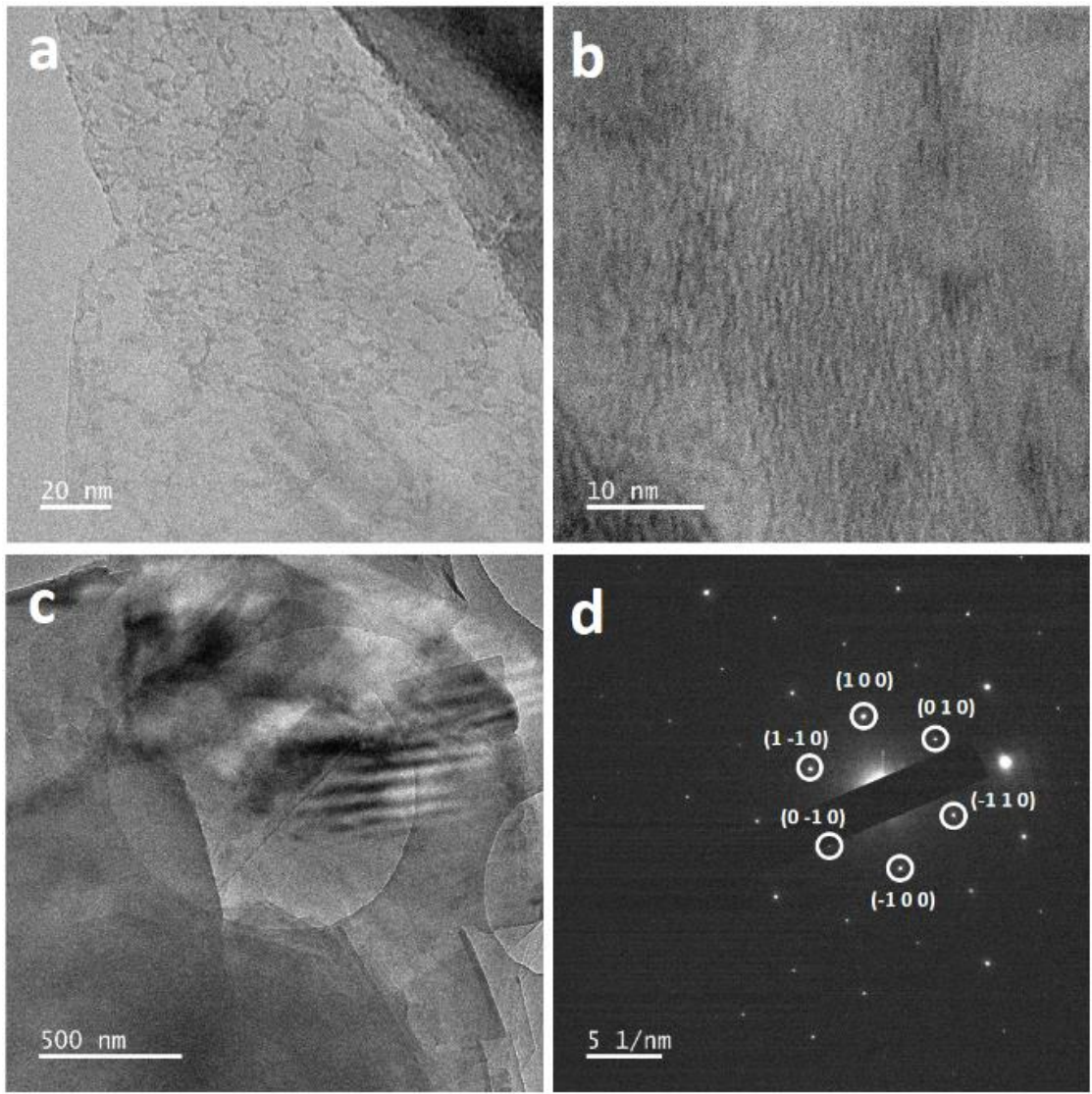

**Figure 10 – TEM of sample after wear test.** a) Worn GNP showing amorphized areas. b) Worn GNP showing a wrinkled morphology. c) A packing of GNPs formed upon wear, thickness fringes are present. d) Indexed diffraction pattern of the worn GNP in (c), showing the [001] zone axis. Here crystallinity is preserved.

A severe degree of structural degradation is shown in Fig. 10a, where a thin, fully amorphized carbon layer is shown. Fig. 10b on the other hand shows a densely packed GNP morphology that can be referred to as wrinkled or rippled [25]. Neither from the area in Fig. 10a nor from that of Fig. 10b had it been possible to obtain crystalline periodicities from the FFT, proving the full amorphization of these areas. None of the particles coming from the as-sprayed sample showed similar morphologies. The samples in Fig. 10a and 10b show morphologies that are compatible with those observed in Fig.8b: the smooth central part, where the highest amount of degradation occurs, and the corrugated sides, respectively.

The TEM image in Fig. 10c shows, at a lower magnification, a packing of GNPs showing a thickness gradient revealed by the contrast fringes on the right side. These are thickness fringes, an interference effect that occurs in thin samples and reveals a thickness variation in it, with equal contrast representing equal thickness [26]. This morphological feature highlights the morphology modifications induced by the wear test. An SAED of this area is shown in Fig. 10d. Here another graphitic single crystal diffraction pattern oriented along the [001] zone axis is shown, similar to that of Fig. 9b, but in this case it is generated by a thicker stack of GNPs, still preserving the crystallinity. The degree of structural modification here is then minimal. However, measuring the distances of the diffraction spots, it has been noticed that along the direction marked as (100) the spots are slightly further from the center than expected. This, converted into real space distances between lattice planes, represents that they are 3% closer along that direction. This observation suggests that, even if the crystallinity is preserved, a very small structural modification is present, which means this area is in an early stage of degradation.

A complementary analysis on a similar sample obtained at the same experimental conditions is presented in [22], where Raman spectroscopy and atomic force microscopy were employed. In that work, it has been concluded that the spray only slightly degrades the GNPs, whereas wear testing causes a noticeable amorphization and structural disorder, both on the single graphene layer and between the stacked graphene layers. The in-flight temperature and velocity measurements provided in the current paper show that GNPs actually undergo very intense heating while being sprayed; however, according to the present results and to the results in [22], this is not enough to destroy them, possibly thanks to the reducing environment of the S-HVOF flame and the very short time they spend in it. The small amount of degradation reported can be caused by the detachment of carbon atoms due to heat, inclusions of oxygen atoms and mechanical shock upon impact with the substrate. The major degradation then occurs upon wear testing, where the GNP structure can in some regions be degraded up to complete amorphization.

## 4 Conclusions

This study focused on radial injection in S-HVOF as a means of depositing heat- and oxygen-sensitive materials. The radial injection dynamics and the feedstock interaction with the S-HVOF jet was observed using high-speed imaging. GNPs were studied as an example of heat- and oxygen-sensitive material, and their morphology, structure and tribology performance were analyzed.

This work showed how radial injection parameters, such as flowrate, injection position and angle, and flame power, affect the feedstock trajectory and vaporization in the S-HVOF flame. Overall, a downstream injection at 10 mm from the gun exit at a moderate flowrate (around 150 ml/min) provides a suitable penetration for a 50 kW flame. The use of Accuraspray 4.0 for S-HVOF allowed to measure to what extent GNP temperature and velocity are varied by the change of these parameters, which affect heat and momentum transfer from the jet to the particles. The temperature and velocity of GNPs has been shown to increase with increasing flame power, decrease with increasing feedstock flowrate and slightly increase with downstream injection.

A set of parameters has been proposed, that provides proper feedstock penetration in the jet, minimizing overspray and potential GNP overheating, transferring enough momentum and heat to obtain a GNP layer while preserving the GNP microstructure and crystallinity. Upon wear testing, the sample exhibits a low coefficient of friction, which stays around 0.1 for 1000 cycles, with an enhancement of up to one order of magnitude compared to the stainless steel case without GNPs. It has been shown that the wear testing can degrade the GNPs forming wrinkled morphology up to causing full amorphization.

This paper provides a broad set of information on radial injection in S-HVOF, which can be useful for a wide range of heat- and oxygen-sensitive materials, opening the way for their effective deposition using this setup.

**Acknowledgments**

This work was supported by the Engineering and Physical Sciences Research Council, Impact Acceleration Account scheme [grant number EP/R511730/1]. The authors gratefully acknowledge the Nanoscale and Microscale Research Centre (nmRC) at the University of Nottingham for access to the SEM and TEM facilities, and Dr. Michael Fay for his help with the TEM experiments. Also, the careful help from R. Screaton and J. Kirk in managing S-HVOF thermal spray is gratefully acknowledged.